\documentclass[12pt]{article}
\usepackage{amsmath,amssymb}
\usepackage[dvips]{graphicx}

\begin{document}

\begin{titlepage}
\begin{flushright}
    OU-HET 440  \\
    hep-th/0305019  \\
    May 2003
\end{flushright}
\begin{center}
  \vspace{3cm}
  {\bf \Large Fuzzy BIon}
  \\  \vspace{2cm}
  Yoshifumi Hyakutake\footnote{E-mail :hyaku@het.sci.phys.osaka-u.ac.jp}
   \\ \vspace{1cm}
   {\it Department of Physics, Osaka University, 
   Toyonaka, Osaka 560-0043, Japan}
\end{center}

\vspace{2cm}
\begin{abstract}
   We construct a solution of the BFSS matrix theory, which is a counterpart 
   of the BIon solution representing a fundamental string ending on a bound state
   of a D2-brane and D0-branes. We call this solution the `fuzzy BIon' and
   show that this configuration preserves 1/4 supersymmetry of type IIA 
   superstring theory.
   We also construct an effective action for the fuzzy BIon by analyzing
   the nonabelian Born-Infeld action for D0-branes. When we take the continuous 
   limit, with some conditions, this action coincides with the effective action for 
   the BIon configuration.
\end{abstract}
\end{titlepage}

\setlength{\baselineskip}{0.65cm}

\vspace{1cm}
\section{Introduction}

It is well recognized that a bound state of D$(p+2)$-brane and D$p$-branes
has dual descriptions. From the D$(p+2)$-brane picture, the D$p$-brane charge
is represented by the magnetic flux on it, and from the D$p$-branes viewpoint,
the D$(p+2)$-brane is realized by some fuzzy configuration of D$p$-branes.
The latter description has the advantage of dealing with multibody systems.
For example it is easy to realize the situation where the D$p$-branes move
around the D$(p+2)$-branes.

In type IIA superstring theory there exists the BIon configuration which
represents a fundamental string ending on a bound state of a D2-brane
and D0-branes\cite{MT}. From the viewpoint of the world-volume theory
on the D2-brane, which is labeled by the radial direction $\rho$ and 
the angular direction $\phi$, 
the fundamental string is expressed as a source of a Coulomb-like electric field 
and D0-branes are realized as uniform magnetic flux on the $(\rho,\phi)$-plane.
One of the transverse scalars, say $z$, on the D2-brane is also excited 
to be the classical solution of field equations.  This configuration 
preserves 1/4 supersymmetry of type IIA superstring theory.

In this paper we give a dual description of the BIon solution, which
is obtained as a classical solution of the BFSS matrix theory.
In order to execute this, we employ the matrix representations 
for the general fuzzy surface with axial symmetry, 
and explicitly write down the equations of motion of the BFSS matrix theory
in terms of their components $\rho_{m+1/2}$, $z_m$ and $a_m$. 
These components contain information
about the shape of a fuzzy surface. 
In fact, $\rho_{m+1/2}$ and $z_m$ almost correspond to the radial 
direction $\rho$ and the transverse scalar $z$ respectively, 
and nontrivial $a_m$ give electric flux on the fuzzy surface. 
We uniquely determine values of $\rho_{m+1/2}$, $z_m$ and 
$a_m$ which equip the properties of the BIon solution. 
We will call this solution the `fuzzy BIon'.

The organization of this paper is as follows.
In section 2 we briefly review the BIon configuration.
In section 3 we solve the equations of motion of BFSS matrix theory and
obtain the solution which represents the fuzzy BIon.  
It will be confirmed that this solution satisfy the
discrete version of the differential equation for the BIon solution and preserves 
1/4 supersymmetry of type IIA superstring theory.
In section 4 the effective action for the fuzzy BIon is constructed by analyzing
the nonabelian Born-Infeld action for D0-branes. When we take the continuous 
limit, with some condition, this action coincides with the effective action
for the BIon configuration. Conclusions are given in section 5.

\section{Review of the BIon solution}

In this section we briefly review the BIon solution which represents
a fundamental string ending on a bound state of a D2-brane and 
D0-branes\cite{MT}.
This brane configuration preserves $1/4$ supersymmetry of type IIA
superstring theory.

Let us choose the line element of the flat space-time as
\begin{alignat}{3}
  ds^2 = -dt^2 + d\rho^2 + \rho^2 d\phi^2 + dz^2 + \sum_{i=4}^9 (dx^i)^2,
\end{alignat}
and identify the world-volume coordinates on the D2-brane with $(t,\rho,\phi)$.
The D2-brane is embedded into the target space like $x^i=0$ and $z=z(\rho)$.
In order to obtain the BIon configuration, we also assume that gauge field strengths
$F_{t\rho}$ and $F_{\rho\phi}$ are nontrivial.
Then the Born-Infeld action for the D2-brane is given by\cite{Lei}
\begin{alignat}{3}
  S_{\text{D2}} = -T_2 \int dtd\rho d\phi \sqrt{\rho^2(1+{z'}^2) + \lambda^2
  F_{\rho\phi}^2 - \rho^2 \lambda^2 F_{t\rho}^2}. \label{eq:act}
\end{alignat}
Here we used the definitions $\lambda = 2\pi\ell_s^2$ and $z'=\frac{dz}{d\rho}$.
$T_2$ is the tension of the D2-brane.

First we consider the world-volume supersymmetry on the D2-brane.
In this case the killing spinor equation $(1-\Gamma) \epsilon =0$ is 
transformed into the form\cite{APS,BT}
\begin{alignat}{3}
  \Big[ \lambda F_{t\rho} \rho \Gamma_{11} \Gamma_{\hat{\phi}} 
  \Big( 1 + \frac{z'}{\lambda F_{t\rho}} \Gamma_{\hat{t}\hat{z}} \Gamma_{11}
  \Big) + \sqrt{X} \Big( 1 - \frac{ \rho \Gamma_{\hat{t} \hat{\rho} \hat{\phi}}
  + \lambda F_{\rho\phi} \Gamma_{\hat{t}} \Gamma_{11} }
  { \sqrt{X} } \Big) \Big] \epsilon = 0,
\end{alignat}
and we defined $X$ as
\begin{alignat}{3}
  X = \rho^2(1+{z'}^2) + \lambda^2 F_{\rho\phi}^2 - \rho^2 \lambda^2 
  F_{t\rho}^2.
\end{alignat} 
The notation $\,\hat{}\;$ for the subscripts of the gamma matrices 
is used to declare the Lorentz indices.
Then the preservation of $1/4$ supersymmetry requires the following
gauge field strength:
\begin{alignat}{3}
  F = \pm \frac{z'}{\lambda} dt \wedge d\rho \pm \frac{\rho}{b} 
  d\rho \wedge d\phi. \label{eq:flu}
\end{alignat}
The first term represents the existence of the constant electric flux 
along the $z$ direction, and plus or minus sign corresponds to the
orientation of the fundamental string.
The second term insists that the magnetic flux projected on the 
$(\rho,\phi)$-plane is uniform, and plus or minus sign corresponds
to the sign of the D0-brane charge.
By using the flux quantization condition of the magnetic flux, we see that 
the area per a unit of magnetic flux projected on the $(\rho,\phi)$-plane 
is $2\pi b$.

Second the Gauss law constraint which is obtained by varying the action with
the gauge field $a_t$ is written as
\begin{alignat}{3}
  \frac{d}{d\rho} \Bigg( \frac{T_2 \lambda^2 F_{t\rho} \rho^2}
  {\sqrt{X}} \Bigg) &= 0.
\end{alignat}
The conditions (\ref{eq:flu}) simplify the above constraint into the form
\begin{alignat}{3}
  dz = \frac{L d\rho^2}{2\rho^2}, \label{eq:BIon}
\end{alignat}
and the solution is written as $z = z_0 + L \ln \rho$.
$L$ and $z_0$ are integral constants.
From this we see that the first term of (\ref{eq:flu}) represents the
Coulomb-like electric field on the $(\rho,\phi)$-plane.
The BIon configuration is characterized by equations (\ref{eq:flu}) and
(\ref{eq:BIon}).

\vspace{0.5cm}
\section{The Fuzzy BIon}

In the previous section we reviewed the BIon configuration by analyzing
the world-volume theory on the D2-brane. Here we
construct a counterpart of the BIon from the viewpoint of D0-branes,
that is, in the framework of the BFSS matrix theory.
We call this solution the `fuzzy BIon'.

\begin{figure}[tb]
\begin{center}
  \includegraphics[width=9cm,height=4cm,keepaspectratio]{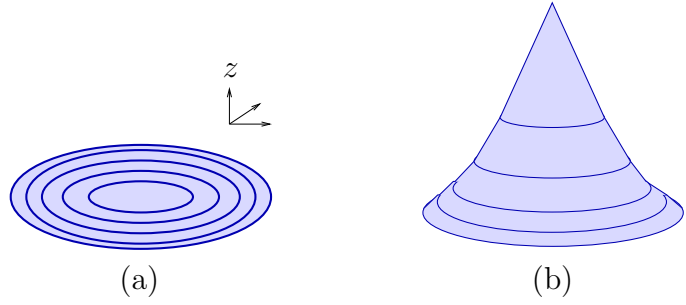}
\begin{picture}(300,0)
  \put(66,0){(a)}
  \put(222,0){(b)}
  \put(105,80){$z$}
\end{picture}
  \vspace{0cm}
  \caption{(a) The fuzzy plane. (b) The fuzzy BIon.
  This configuration can be obtained by pulling the fuzzy plane 
  into the $z$ direction.}
  \label{fig:fBIon}
\end{center}
\end{figure}

Let us begin with the action of the BFSS Matrix Theory\cite{BFSS},
\begin{alignat}{3}
  S_{\text{BFSS}} &= T_0 \int dt \text{Tr} \Big( \frac{1}{2} (D_t X^i)^2
  + \frac{1}{4\lambda^2} [X^i,X^j]^2 \Big),
\end{alignat}
where $i,j$ run $1,\cdots,9$.
The covariant derivative is defined as $D_t X^i = \partial_t X^i + i[A_t,X^i]$,
and $T_0$ is the mass of a D0-brane.
The equations of motion for $X^i$ and the Gauss law constraint on $A_t$ are
written as
\begin{alignat}{3}
  -D_t (D_t X^i) + \frac{1}{\lambda^2} [X^j,[X^i,X^j]] = 0, \qquad
  [X^i , D_t X^i] = 0. \label{eq:eom}
\end{alignat}
Now we solve these equations to obtain the fuzzy BIon solution.
In order to execute this, we set $X^4 = \cdots = X^9 = 0$ and choose the 
matrices $X^1$, $X^2$, $X^3$ and $A_t$ like
\begin{alignat}{3}
  X^1_{mn} &= \frac{1}{2} \rho_{m+1/2} \delta_{m+1,n} + \frac{1}{2}
  \rho_{m-1/2} \delta_{m,n+1}, \notag
  \\
  X^2_{mn} &= \frac{i}{2} \rho_{m+1/2} \delta_{m+1,n} - \frac{i}{2}
  \rho_{m-1/2} \delta_{m,n+1}, 
  \\[0.2cm]
  X^3_{mn} &= z_m \delta_{m,n} ,\qquad A_{t\, mn} = a_{m} 
  \delta_{m,n}, \notag
\end{alignat}
where $m,n \in \mathbb{N}$. These matrices represent the fuzzy surface 
with axial symmetry around the $x^3 \,(= z)$ direction\cite{Hya1,Hya2}.
Since what we want is a static configuration, each component does not depend 
on the time $t$. After some calculations, the equations (\ref{eq:eom}) are 
transformed into the forms,
\begin{alignat}{3}
  &2\lambda^2(a_{m+1}-a_{m})^2 - 2(z_{m+1}-z_m)^2
  + (\rho_{m+3/2}^2 - 2\rho_{m+1/2}^2 + \rho_{m-1/2}^2) = 0, \notag
  \\[0.1cm]
  &\rho_{m+1/2}^2 (z_{m+1}-z_m) - \rho_{m-1/2}^2 (z_m-z_{m-1}) = 0,
  \label{eq:eom2}
  \\[0.1cm]
  &\rho_{m+1/2}^2 (a_{m+1}-a_{m}) - \rho_{m-1/2}^2 
  (a_{m}-a_{m-1}) = 0. \notag
\end{alignat}
Here we should set $\rho_{1/2} = 0$. Note that when
$z_m$ and $a_m$ are all equal to zero, we obtain nontrivial solutions
by choosing $\rho_{m+1/2}$ as
\begin{alignat}{3}
  \rho^c_{m+1/2} = \sqrt{2bm}. \label{eq:rho}
\end{alignat}
The superscript $c$ is denoted to emphasize that this is a classical solution.
These matrices are known to represent 
the fuzzy plane\cite{BFSS,Hya2}. (see Fig.\,\ref{fig:fBIon} (a).)
The commutation relation for $X^1$ and $X^2$ becomes
$[X^1,X^2] = -ib$, and $2\pi b$ represents the fuzziness of a D0-brane.

The explicit values of $\rho_{m+1/2}$, $z_m$ and $a_m$ which represent
the fuzzy BIon can be obtained by referring the equation (\ref{eq:flu}) obtained 
in the previous section. The latter term of (\ref{eq:flu}) insists that 
the magnetic flux of the BIon configuration is uniform on the projected 
$(\rho,\phi)$-plane and the area occupied per a unit of magnetic flux is $2\pi b$.
If we neglect the electric flux,
this means that the BIon configuration can be obtained by pulling the 
planar D2-brane with uniform magnetic flux on it to the $z$ direction.
Of course, the equations of motion determined the form of the function $z(\rho)$ 
and the preservation of 1/4 supersymmetry required the existence of 
the electric flux.

Now we trace the same procedure as the above.
First we identify $\rho_{m+1/2}$ for the fuzzy BIon configuration with
that of the fuzzy plane. Then by substituting $\rho_{m+1/2} = \sqrt{2bm}$ into 
(\ref{eq:eom2}), the equations for $z_m$ and $a_m$ are obtained like
\begin{alignat}{3}
  &z_{m+1}-z_m = \pm\lambda(a_{m+1}-a_{m}), \notag
  \\[0.1cm]
  &m (z_{m+1}-z_m) - (m-1) (z_m-z_{m-1}) = 0,
  \label{eq:eom3}
\end{alignat}
and $z_m$ and $a_m$ are easily solved as
\begin{alignat}{3}
  z^c_m = \pm \lambda a^c_m 
  = z^c_1 + \frac{L}{2} \sum_{i=1}^{m-1} \frac{1}{i}. \label{eq:z}
\end{alignat}
Here $m \geq 2$ and $L$ and $z^c_1 = \pm \lambda a^c_1$ are some constant 
parameters.
This solution insists the existence of fundamental string charge 
along the $z$ direction because of the relation
$A_t = \pm \frac{1}{\lambda} X^3$\cite{BSS,BL,Hya3}. 
The sign corresponds to the orientation of the fundamental string.
We can confirm that this solution really represents the fuzzy 
BIon configuration, by noting the relation
\begin{alignat}{3}
  z^c_{m+1} - z^c_{m} = \frac{L ( \rho_{m+1}^{c\,2} - \rho_{m}^{c\,2}) }
  {2 \rho_{m+1/2}^{c\,2}}, \label{eq:fBIon}
\end{alignat}
where $\rho_{m+1}^{c\,2} \equiv (\rho_{m+3/2}^{c\,2} 
+ \rho_{m+1/2}^{c\,2})/2$.
This is the matrix version of the equation (\ref{eq:BIon}).

The commutation relations which represents the fuzzy BIon are written as
\begin{alignat}{3}
  &[X^1,X^2]_{mn} = -ib \delta_{m,n}, \notag
  \\[0.1cm]
  &[X^2,X^3]_{mn} = \frac{i}{2} bL\rho_{m+1/2}^{c\;-1}  \delta_{m+1,n}
  + \frac{i}{2} bL\rho_{m-1/2}^{c\;-1} \delta_{m,n+1}, 
  \\
  &[X^3,X^1]_{mn} = - \frac{1}{2} bL\rho_{m+1/2}^{c\;-1} \delta_{m+1,n}
  + \frac{1}{2} bL\rho_{m-1/2}^{c\;-1} \delta_{m,n+1}. \notag
\end{alignat}
And the figure of the fuzzy BIon is drawn in Fig.\,\ref{fig:fBIon} (b).

Our final confirmation is to check the preservation of 1/4 supersymmetry.
In this case the killing spinor equation $\delta\Theta = 0$ becomes
\begin{alignat}{3}
  0 &= \Big( D_t X^i \Gamma^{0i} + \frac{i}{2\lambda}
  [X^i,X^j] \Gamma^{ij} \Big) P_+ \epsilon 
  + P_+ \epsilon' 
  \\
  &= \pm \frac{i}{\lambda} \big( [X^3,X^1] \Gamma^{01} + [X^3,X^2] 
  \Gamma^{02} \big) P_+
  \big(1 \mp \Gamma^{03} \Gamma_{11} \big) \epsilon 
  + \frac{i}{\lambda} [X^1,X^2] P_+ \Gamma^{12} \epsilon + P_+ \epsilon', \notag
\end{alignat}
where $\epsilon$ and $\epsilon'$ are constant Majorana spinors and
$P_+ = (1+\Gamma_{11})/2$.
Thus we see that 1/4 supersymmetry is preserved when
\begin{alignat}{3}
  \epsilon' = \frac{b}{\lambda} \Gamma^0 \epsilon , \qquad
  \epsilon = \frac{1\pm\Gamma^{03}\Gamma_{11}}{2}
  \frac{1+\Gamma^{012}}{2} \epsilon_0, 
\end{alignat}
where $\epsilon_0$ is an arbitrary constant Majorana spinor.

\vspace{0.5cm}
\section{The Effective Action for the Fuzzy BIon}

In the previous section we obtained the fuzzy BIon configuration as the classical
solution of the BFSS matrix theory. We also checked that the fuzzy BIon 
preserves 1/4 supersymmetry of type IIA superstring theory. 
Now it is natural to ask whether the effective action for D0-branes
which is described by the nonabelian Born-Infeld action
contains the same solution. In this section we do not prove it directly, but
construct the effective action for the fuzzy BIon by analyzing the nonabelian 
Born-Infeld action for D0-branes. 
We show that, with some conditions, the effective action obtained in this way 
correctly reproduce that for the BIon configuration in the continuous limit.

The nonabelian Born-Infeld action is given by refs.\,\cite{Tse,Tse2,Mye}.
In the background of the flat space-time, when we set  $X^4 = \cdots = X^9 = 0$,
this action is transformed into the form\cite{Hya2}
\begin{alignat}{3}
  S_{\text{D}0} &= -T_0 \!\int\! dt \text{Tr} \sqrt{ 1 - (D_t X^i)^2
  - \frac{1}{2\lambda^2}[X^i,X^j]^2 + \frac{1}{4 \lambda^2}
  \big( \epsilon_{ijk} \big\{D_t X^i, [X^j,X^k] \big\} \big)^2 }, \label{eq:D0}
\end{alignat}
where $i,j,k = 1,2,3$ and $\{A,B\} \equiv (AB+BA)/2$.
Now we choose representations of three adjoint scalars $X^1$, $X^2$, $X^3$ and
a gauge field $A_t$ as
\begin{alignat}{3}
  X^1_{mn} &= \frac{1}{2} \rho e^{i\theta} \big|_{m+1/2} \delta_{m+1,n} + 
  \frac{1}{2} \rho e^{-i\theta} \big|_{m-1/2} \delta_{m,n+1}, \notag
  \\
  X^2_{mn} &= \frac{i}{2} \rho e^{i\theta} \big|_{m+1/2} \delta_{m+1,n} -
  \frac{i}{2} \rho e^{-i\theta} \big|_{m-1/2} \delta_{m,n+1}, 
  \\[0.2cm]
  X^3_{mn} &= z |_m \delta_{m,n} ,\qquad 
  A_{t\, mn} = a |_m \delta_{m,n}. \notag
\end{alignat}
The notation $|_{m}$ is employed to show functions before this have
the subscript $m$. Here the elements $\rho_{m+1/2}$, $\theta_{m+1/2}$, $z_m$
and $a_m$ are functions of  the time $t$. Then by introducing the 
following definitions
\begin{alignat}{3}
  &\delta z|_{m+1/2} = z|_{m+1} - z|_m , \notag
  \\[0.2cm]
  &\delta a|_{m+1/2} = a|_{m+1} - a|_m, \notag
  \\[0.2cm]
  &\rho^2\delta z^2|_m = \frac{1}{2} \big( \rho^2\delta z^2|_{m+1/2}
  + \rho^2 \delta z^2|_{m-1/2} \big) ,\notag
  \\[0.2cm]
  &\delta \rho^2|_{m} = \rho^2|_{m+1/2} - \rho^2|_{m-1/2}, \label{eq:defs}
  \\[0.2cm]
  &\dot{\rho}^2|_{m} = \frac{1}{2} \big( \dot{\rho}^2|_{m+1/2} 
  + \dot{\rho}^2|_{m-1/2} \big) ,  \notag
  \\
  &\rho^2 (\dot{\theta} - \delta a)^2|_m = \frac{1}{2} 
  \big\{ \rho^2 (\dot{\theta} - \delta a)^2|_{m+1/2} 
  + \rho^2 (\dot{\theta} - \delta a)^2|_{m-1/2} \big\}, \notag
  \\
  &\rho\dot{\rho}\delta z|_m = \frac{1}{2} \big( \rho\dot{\rho}\delta z|_{m+1/2}
  + \rho\dot{\rho}\delta z|_{m-1/2} \big) , \notag
\end{alignat}
after some calculations, the effective action (\ref{eq:D0}) is evaluated as
\begin{alignat}{3}
  S_{\text{D}0} &= -T_0 \!\int\! dt \sum_m \bigg[ 1 + \Big\{ - \dot{z}^2 
  + \frac{1}{\lambda^2} \rho^2\delta z^2 \Big\} \notag
  \\
  &\qquad\qquad\qquad\qquad
  + \lambda^2 \Big\{ -\frac{1}{\lambda^2} \rho^2 (\dot{\theta} - \delta a)^2 
  + \frac{1}{4\lambda^4} (\delta \rho^2)^2 
  - \frac{1}{\lambda^2} \dot{\rho}^2 \Big\}  \label{eq:D02}
  \\[0.1cm]
  &\qquad\qquad\qquad\qquad
  - \lambda^2 \Big\{ \frac{1}{2\lambda^2} \dot{z} \delta\rho^2 
  - \frac{1}{\lambda^2} \rho\dot{\rho}\delta z \Big\}^2 \bigg]^{1/2} \bigg|_m .
  \notag
\end{alignat}
Note that the interior of the square root is proportional to the identity matrix and
the trace operation is simply replaced with the sum.

Let us return to the case of the fuzzy BIon configuration. 
Now we add a scalar fluctuation $\hat{z}|_m$ and gauge fluctuations
$a_\rho|_{m+1/2}$, $a_\phi|_{m+1/2}$ and $a_t|_m$ around the fuzzy BIon
configuration (\ref{eq:rho}) and (\ref{eq:z}) like
\begin{alignat}{3}
  X^1_{mn} &= \frac{1}{2} \big( \rho^c + la_{\phi} \big) 
  e^{ila_{\rho}} \big|_{m+1/2} \delta_{m+1,n} 
  + \frac{1}{2} \big( \rho^c + la_{\phi} \big) 
  e^{-ila_{\rho}} \big|_{m-1/2} \delta_{m,n+1}, \notag
  \\
  X^2_{mn} &= \frac{i}{2} \big( \rho^c + la_{\phi} \big) 
  e^{ila_{\rho}} \big|_{m+1/2} \delta_{m+1,n} 
  - \tfrac{i}{2} \big( \rho^c + la_{\phi} \big) 
  e^{-ila_{\rho}} \big|_{m-1/2} \delta_{m,n+1}, 
  \\[0.2cm]
  X^3_{mn} &= \big( z^c + \hat{z} \big) \big|_m \delta_{m,n} , \qquad
  A_{t\,mn} = \big( a^c + a_t \big) \big|_m \delta_{m,n}. \notag
\end{alignat}
Here we defined $l|_{m+1/2} = b/\rho^c|_{m+1/2}$ which is interpreted
as a separation between $m$th and $(m+1)$th segments. 
We also define $\rho^c|_m = (\rho^c|_{m+1/2} + \rho^c|_{m-1/2})/2$ 
and $l|_m = b/\rho^c|_m$
for later use. Now the definitions (\ref{eq:defs}) are translated into the forms
\begin{alignat}{3}
  &\delta z|_{m+1/2} = \frac{b}{\rho^c} \Big( \frac{L}{\rho^c}
  \!+\! \hat{z}' \Big) \Big|_{m+1/2} , 
  \qquad\qquad\;\,
  \hat{z}'|_{m+1/2} = \frac{\hat{z}|_{m+1} - \hat{z}|_m}{l|_{m+1/2}}, \notag
  \\[0.1cm]
  &\delta a|_{m+1/2} = \frac{b}{\rho^c} \Big( \!\!\pm\! \frac{L}{\lambda\rho^c}
  \!+\! a'_t \Big) \Big|_{m+1/2} , 
  \qquad\quad
  a'_t|_{m+1/2} = \frac{a_t|_{m+1} - a_t|_m}{l|_{m+1/2}}, \notag
  \\[0.1cm]
  &\delta \rho^2|_{m} \cong \frac{2b^2}{\rho^c} 
  \Big( \frac{\rho^c}{b} \!+\! a'_\phi \Big) \Big|_m, 
  \qquad\qquad\qquad\;\;
  a'_\phi|_m = \frac{a_\phi|_{m+1/2} - a_\phi|_{m-1/2}}{l|_m}, \notag
  \\[0.1cm]
  &\rho^2\delta z^2|_m \cong
  \frac{1}{2} \Big\{ b^2 \Big( \frac{L}{\rho^c} \!+\! \hat{z}' \Big)^2 \Big|_{m+1/2}
  \!\!+ b^2 \Big( \frac{L}{\rho^c} \!+\! \hat{z}' \Big)^2 \Big|_{m-1/2} \Big\}
  \equiv b^2 \Big( \frac{L}{\rho^c} \!+\! \hat{z}' \Big)^2 \Big|_m , \label{eq:defs2}
  \\[0.1cm]
  &\dot{\rho}^2|_{m} = \frac{1}{2} \Big( \frac{b^2}{\rho^{c\,2}} 
  \dot{a}_{\phi}^2 \Big|_{m+1/2} 
  \!\!+ \frac{b^2}{\rho^{c\,2}} \dot{a}_{\phi}^2 \Big|_{m-1/2} \Big)
  \equiv \frac{b^2}{\rho^{c\,2}} \dot{a}_{\phi}^2 \Big|_m ,  \notag
  \\[0.1cm]
  &\rho^2 (\dot{\theta} \!-\! \delta a)^2|_m \cong \frac{1}{2} 
  \Big\{ b^2 \Big( \!\!\mp\! \frac{L}{\lambda\rho^c} \!+\! 
  f_{t\rho} \Big)^2 \Big|_{m+1/2}
  \!\!+ b^2 \Big( \!\!\mp\! \frac{L}{\lambda\rho^c} \!+\! f_{t\rho} \Big)^2 
  \Big|_{m-1/2} \Big\} 
  \equiv b^2 \Big( \!\!\mp\! \frac{L}{\lambda\rho^c} \!+\! 
  f_{t\rho} \Big)^2 \Big|_m , \notag
  \\[0.1cm]
  &\rho\dot{\rho}\delta z|_m \cong \frac{1}{2} \Big\{ \frac{b^2}{\rho^c} 
  \Big( \frac{L}{\rho^c} \!+\! \hat{z}' \Big) \dot{a}_\phi \Big|_{m+1/2}
  \!\!+ \frac{b^2}{\rho^c} \Big( \frac{L}{\rho^c} \!+\! \hat{z}' \Big) \dot{a}_\phi
  \Big|_{m-1/2} \Big\} 
  \equiv \frac{b^2}{\rho^c} \Big( \frac{L}{\rho^c} \!+\! \hat{z}' \Big) 
  \dot{a}_\phi \Big|_m . \notag
\end{alignat}
The symbol $\cong$ is used when we ignore higher order terms 
on $l a_\phi/\rho^c|_m$.
By substituting these values into (\ref{eq:D02}), 
we obtain the effective action for the fuzzy BIon configuration,
\begin{alignat}{3}
  S_{\text{D}0} &\cong -T_0 \!\int\! dt \sum_m \frac{l\rho^c}{b} 
  \bigg[ 1 + \Big\{ - \dot{\hat{z}}^2 + \frac{b^2}{\lambda^2} 
  \Big( \frac{L}{\rho^c} + \hat{z}' \Big)^2 \Big\} \notag
  \\
  &\qquad\qquad
  + \lambda^2 \Big\{ -\frac{b^2}{\lambda^2} 
  \Big( \mp \frac{L}{\lambda\rho^c} + f_{t\rho} \Big)^2
  + \frac{b^4}{\lambda^4 \rho^{c\,2}} 
  \Big( \frac{\rho^c}{b} + a'_\phi \Big)^2
  - \frac{b^2}{\lambda^2\rho^{c\,2}} \dot{a}_\phi^2 \Big\}  \label{eq:D03}
  \\
  &\qquad\qquad
  - \lambda^2 \Big\{ \frac{b^2}{\lambda^2\rho^c} \dot{\hat{z}}
  \Big( \frac{\rho^c}{b} + a'_\phi \Big)
  - \frac{b^2}{\lambda^2\rho^c} \Big( \frac{L}{\rho^c} + \hat{z}' \Big) 
  \dot{a}_\phi \Big\}^2 \bigg]^{1/2} \bigg|_m . \notag
\end{alignat}
Note that we used the relations $\rho^c l|_{m+1/2} = \rho^c l|_m = b$.
Let us consider that $l_m$ are sufficiently small and 
take the continuous limit. Then the above effective action for the fuzzy BIon 
reaches to
\begin{alignat}{3}
  S_{\text{D}0} &= -\frac{T_0}{b} \!\int\! dt d\rho \rho
  \bigg[ 1 + \Big\{ - \dot{\hat{z}}^2 + \frac{b^2}{\lambda^2} 
  \Big( \frac{L}{\rho} + \hat{z}' \Big)^2 \Big\} \notag
  \\
  &\qquad\qquad
  + \lambda^2 \Big\{ -\frac{b^2}{\lambda^2} 
  \Big( \mp \frac{L}{\lambda\rho} + f_{t\rho} \Big)^2
  + \frac{b^4}{\lambda^4 \rho^2} 
  \Big( \frac{\rho}{b} + a'_\phi \Big)^2
  - \frac{b^2}{\lambda^2\rho^2} \dot{a}_\phi^2 \Big\}  \label{eq:D04}
  \\
  &\qquad\qquad
  - \lambda^2 \Big\{ \frac{b^2}{\lambda^2\rho} \dot{\hat{z}}
  \Big( \frac{\rho}{b} + a'_\phi \Big)
  - \frac{b^2}{\lambda^2\rho} \Big( \frac{L}{\rho} + \hat{z}' \Big) 
  \dot{a}_\phi \Big\}^2 \bigg]^{1/2}  . \notag
\end{alignat}
In order to justify this action, we should compare with the effective action 
for the BIon configuration.

The effective action for the D2-brane is described by
\begin{alignat}{3}
  S_{\text{D2}} = - T_2 \int dtd\rho d\phi \rho \sqrt{1 + \partial_\alpha z
  \partial^\alpha z + \frac{\lambda^2}{2} F_{\alpha\beta} F^{\alpha\beta}
  - \frac{\lambda^2}{4} (\epsilon^{\alpha\beta\gamma} \partial_\alpha z
  F_{\beta\gamma})^2 },
\end{alignat}
where indices are raised or lowered by the metric
$ds^2 = -dt^2 + d\rho^2 + \rho^2 d\phi^2$ and 
$\epsilon^{t\rho\phi} = 1/\rho$.
The fluctuations around the BIon solution is given by
\begin{alignat}{3}
  &z = z_0 + L \ln \rho + \hat{z}(t,\rho), \notag
  \\[0.1cm]
  &F_{t\rho} = \pm \frac{L}{\lambda\rho} + f_{t\rho}(t,\rho),
  \\
  &F_{\rho\phi} = \frac{\rho}{b} + a'_\phi(t,\rho), \notag
  \\[0.2cm]
  &F_{t\phi} = \dot{a}_\phi(t,\rho). \notag
\end{alignat}
We assumed that fluctuations $\hat{z}$, $a_t$, $a_\rho$
and $a_\phi$ do not depend on the angular direction $\phi$.
By substituting these into the effective action, we obtain
\begin{alignat}{3}
  S_{\text{D2}} &= - 2\pi T_2 \int dtd\rho \rho \bigg[ 1 + 
  \Big\{ -\dot{\hat{z}}^2 + \Big(\frac{L}{\rho} + \hat{z}'\Big)^2 \Big\} \notag
  \\[0.2cm]
  &\qquad\qquad\qquad\qquad\;
  + \lambda^2 \Big\{ -\Big( \mp \frac{L}{\lambda\rho} + f_{t\rho} \Big)^2 
  + \frac{1}{\rho^2} \Big( \frac{\rho}{b} + a'_\phi \Big)^2
  - \frac{1}{\rho^2} {\dot{a}_\phi}^2 \Big\} \label{eq:D2}
  \\[0.1cm]
  &\qquad\qquad\qquad\qquad\;
  - \lambda^2 \Big\{ \frac{1}{\rho} \dot{\hat{z}} \Big( \frac{\rho}{b} 
  + a'_\phi \Big) - \frac{1}{\rho} \Big( \frac{L}{\rho} + \hat{z}' \Big) 
  \dot{a}_\phi \Big\}^2 \bigg]^{1/2} . \notag
\end{alignat}
From these we see that the action (\ref{eq:D04}) coincides with the action 
(\ref{eq:D2}) in the case of $b=\lambda$.
This result is the same as that in ref.\,\cite{Hya2}.
The condition $b=\lambda$ suggests that a D0-brane can transform into
a D2-brane with the area $2\pi\lambda$.

\vspace{0.5cm}
\section{Conclusion}

In this paper we construct the fuzzy BIon configuration in the framework of
BFSS matrix theory. This solution equips the properties of 
the BIon solution which represents a fundamental string
ending on a bound state of a D2-brane and D0-branes.
For instance, the differential equation (\ref{eq:BIon}) for the BIon
corresponds to the discrete equations (\ref{eq:fBIon}) for the fuzzy BIon.
We also checked that the fuzzy BIon preserves 1/4 supersymmetry of
type IIA superstring theory.

In section 4, we construct the effective action for the fuzzy BIon configuration
by analyzing the nonabelian Born-Infeld action for D0-branes.
In the continuous limit, this action coincides with the effective 
action for the BIon configuration, in the case of $b = \lambda$.
This means that a D0-brane can transform into a 
D2-brane with the area $2\pi \lambda$.
This fact is also supported by the energy conservation that
$T_0 = 2\pi \lambda T_2$.
With these nontrivial confirmations, we conclude that the fuzzy BIon configuration
is also a solution of nonabelian Born-Infeld action for D0-branes.

It is important to note that only a fundamental string exists 
around the origin of the fuzzy BIon. This would make us possible 
to compute the interaction between a fundamental string and D0-branes
in the framework of BFSS matrix theory.
It is also interesting to generalize the fuzzy BIon configuration
in the curved space background\cite{PTM}.

\vspace{0.5cm}
\section*{Acknowledgements}

\vspace{0.3cm}
I would like to thank Tsuguhiko Asakawa, So Matsuura and Nobuyoshi Ohta.
The work was supported in part by the Grant-in-Aid for JSPS fellows.

\vspace{1cm}
\noindent
\textbf{Note added:} 
In ref.\,\cite{NK} the same equations as (\ref{eq:eom2}) are obtained. 
I would like to thank Nakwoo Kim for pointing out this fact.

\end{document}